\documentclass[a4paper,final]{appolb}
\usepackage{epsfig}

\begin{document}
 \eqsec  
\title{Directed flow of Identified Charged Particles from the RHIC Beam Energy Scan  
\thanks {Presented at the conference ``Strangeness in Quark Matter 2011", Cracow, Poland}}
\author{Yadav Pandit (for the STAR Collaboration) 
\address{Kent State University, Kent, OH 44240}}
\maketitle
\begin{abstract}
We  present  the STAR measurements of directed flow, $v_1$,  for $\pi^\pm$, $K^\pm$, protons and antiprotons, as well  as for all detected charged particles in Au + Au collisions at $\sqrt{s_{NN}} = $ 7.7, 11.5 and 39 GeV as a function of transverse momentum, rapidity and centrality. Results are compared to the predictions from transport models.
\end{abstract}
  
\section{Introduction}
The RHIC Beam Energy Scan (BES) is mainly focused on a search for the QCD critical point and the 1st-order phase boundary in the QCD phase diagram. In the year 2010, the STAR experiment at RHIC took data for the beam energies $\sqrt{s_{NN}} = $7.7, 11.5 and 39 GeV  as a first phase of the BES program.  Analysis of the collected data for various observables sensitive to the phase transition and critical point  continues for the successful accomplishment of the program.  

Directed flow, $v_1$, is the first harmonic coefficient in the azimuthal distribution of the particles with respect to the reaction plane,
\begin{equation}
v_1 = \langle \cos ( \phi-\Psi_R ) \rangle   ,
\end{equation}
where $\phi$ denotes the azimuthal angle of an outgoing particle and  $\Psi_R$ is the orientation 
of the reaction plane defined by the beam axis and the impact parameter vector\cite{methods}. Both hydrodynamic and nuclear transport models~\cite{Hydro,Transport} indicate that directed flow is a sensitive signature for the phenomena related to the possible phase transition specially in the BES region~\cite{bes}. In particular, the shape of $v_1(y)$ may exhibit flatness at midrapidity due to a strong, tilted expansion of the source giving rise to anti-flow  or a 3rd flow~\cite{Csernai}  component.  If the tilted expansion is strong enough, it can cancel and reverse the motion in the bounce-off  direction and results in a negative $v_{1}(y)$ slope at midrapidity, potentially producing a wiggle like structure in $v_1(y)$.  A wiggle for baryons is a possible signature of a phase transition between hadronic matter and Quark Gluon Plasma (QGP), although QGP is not the only possible explanation~\cite{Csernai, Brachmann, Stocker}.  If strong but incomplete baryon stopping is  assumed together with positive space-momentum correlations caused by transverse radial expansion, then a wiggle structure might be explained even in a hadronic system~\cite{Wiggle}.  
\section{Methods and Analysis}
 \begin{figure}
\begin{center}
\includegraphics[scale=0.59]{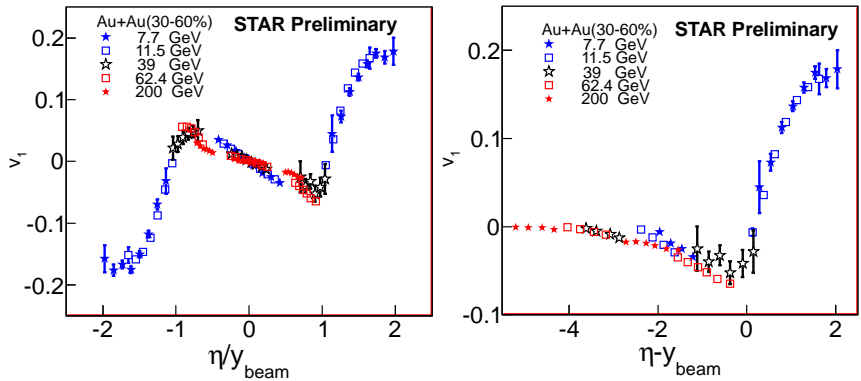}
\caption{ The left panel shows charged hadron $v_1$ as a function of $\eta$ scaled by  $y_{\rm{beam}}$  for Au + Au collisions at $\sqrt{s_{NN}}$ = 7.7, 11.5, 39, 62.4 and 200 GeV  for 30--60\% central collisions. The right panel shows charged hadron $ v_{1} $ as a function of $\eta$ -  $y_{\rm{beam}}$.  The results for 62.4 and 200 GeV are for 30--60\% centrality, previously reported by STAR~\cite{v1-4systems}.}
\label{fig1}
\end{center}
\end{figure}

\begin{figure} 
\begin{center}
\includegraphics[scale=0.63]{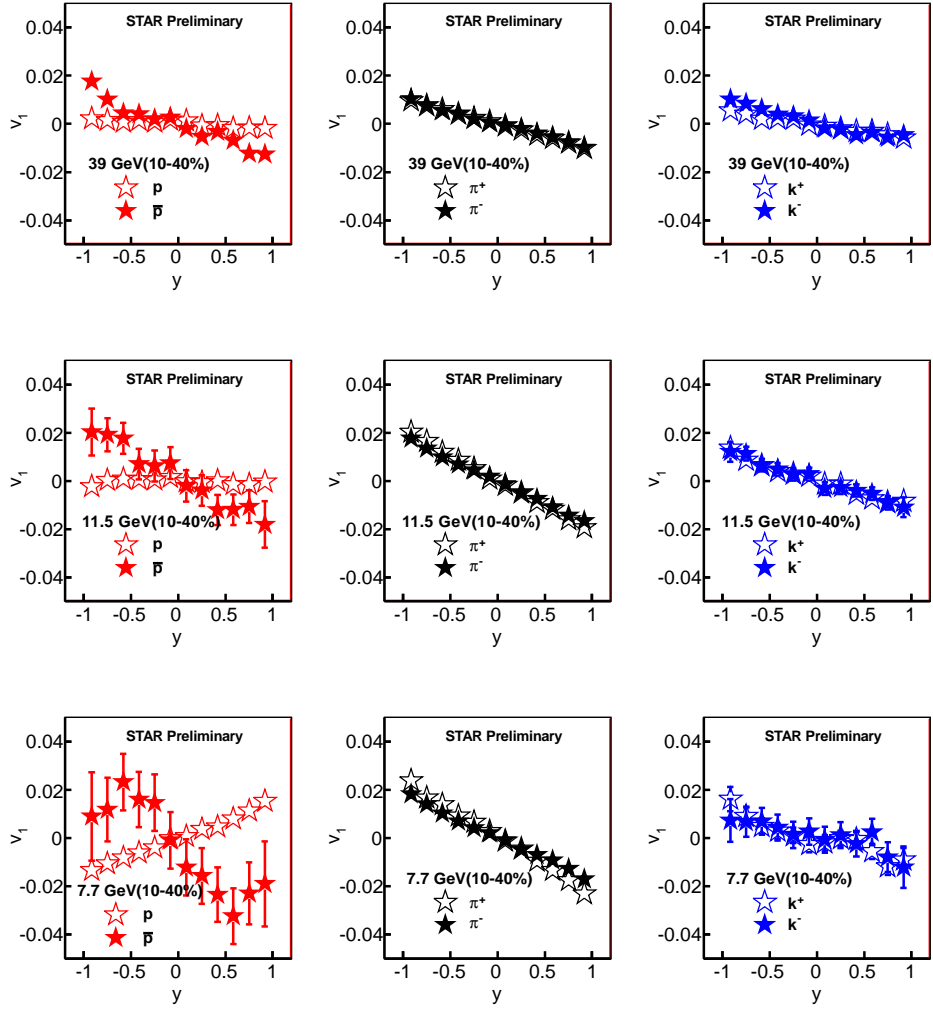}
\caption{The top, middle and bottom row of panels  show directed flow $v_1$ as a function of rapidity for positive and negative hadrons from Au+Au collisions at $\sqrt{s_{NN}} =$  7.7, 11.5 and 39 GeV respectively with proton, antiproton, $\pi^{\pm}$  and $K^{\pm} $ shown in the  first, second and  last  column  of panels  for mid-central (10-40\%) collisions. }
\label{fig2}
\end{center}
\end{figure}

 In these  proceedings, we report $ v_{1} $   measurements by the STAR experiment from $\sqrt{s_{NN}}$  = 7.7, 11.5 and  39 GeV Au + Au  collisions. Data were taken from Run 10 (2010). The STAR Time Projection Chamber (TPC) \cite{startpc} was used as the main detector for charged particle tracking at midrapidity  and Forward Time Projection Chambers (FTPCs) were  used for charged particle  tracking at forward rapidities.  The centrality was determined by the number of tracks from the pseudorapidity region $ |\eta|  < 0.5 $.  Two  Beam Beam Counters  covering  $3.3  < |\eta|  < 5.0$ were used to reconstruct the first order  event plane for this  analysis. The pseudorapidity gap between BBC and TPC suppresses the the non-flow effects.  We analyzed minimum bias events with the following conditions: event vertex radius ($\sqrt{V_x^2 + V_y^2}) < 2.0$ cm, where $V_x$ and $V_y$ are the vertex positions along the $x$ and $y$ directions, respectively, and $V_z$ within 70, 50 and 30 cm of the center of the detector at  7.7, 11.5 and  39 GeV respectively. Tracks are also required to have transverse momenta $ p_T > 0.2 $ GeV$/c$, pass within 3 cm of the primary vertex, have at least 15 space points in the main TPC acceptance  $(|\eta| < 1.0)$ or 5 space points in the case of tracks in the FTPC acceptance ($2.5 < |\eta| < 4.0$), and we require the ratio of the number of actual space points to the maximum possible number of space points to be greater than 0.52.  Protons and antiprotons up to 2.8 GeV/$c$ and  $\pi^\pm$ and $K^\pm$  up to 1.6 GeV/$c$ in transverse momentum  were identified based on specific energy loss in the TPC and the time-of-flight information from multi-gap resistive plate chamber TOF detectors in combination with the momentum.  
 \section{Results} 
 
Beam energy and centrality dependence of directed flow of charged particles as a function of pseudorapidity and transverse momentum has been previously reported \cite{ypflow}. Fig~\ref{fig1} (left panel) shows charged hadron $v_1$ as a function of $\eta$ scaled by the corresponding beam rapidity($y_{\rm{beam}}$) and right panel as a function of $\eta$ - $y_{\rm{beam}}$ for Au+Au Collisions at $\sqrt{s_{NN}}$=7.7, 11.5, 39, 62.4 and 200 GeV  for 30--60\% central collisions. The new results reported here are the charged hadron $v_1\{{\rm BBC}\}$ in Au+Au collisions for 30--60\% centrality at $\sqrt{s_{NN}} = $ 7.7, 11.5 and 39 GeV. We observe  an approximate beam energy scaling behavior of directed flow~\cite{CuCu22}, also observed at SPS energies~\cite{NA49}.
 
Fig.~\ref{fig2} shows directed flow $v_{1}$ of identified particles proton, antiproton, $\pi^{\pm}$  and $K^{\pm} $ as a function of rapidity for  from Au+Au collisions at $\sqrt{s_{NN}} $ =  7.7, 11.5 and 39 GeV. We observe the differences in $v_{1}$ for  protons and antiprotons. 
 \begin{figure}
\begin{center}
\includegraphics[scale=0.63]{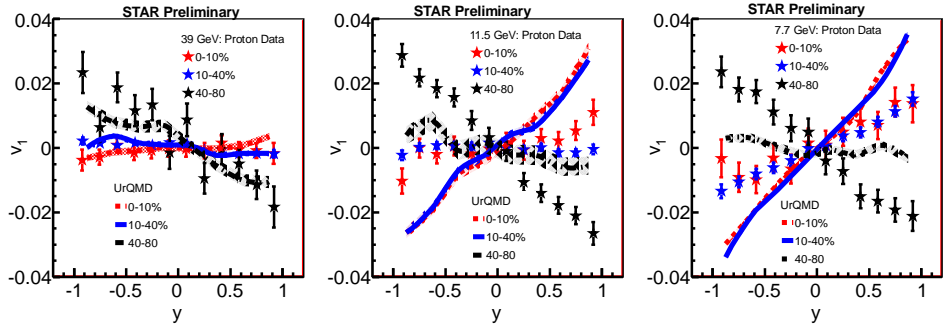}
\caption{Centrality dependence of directed flow of protons as a  function of rapidity, y. Results are compared with the UrQMD model calculations.} 
\label{fig3}
\end{center}
\end{figure}

In Fig.~\ref{fig3}, proton $v_1$ as a function of rapidity ($y$) for central (0-10\%), midcentral (10-40\%) and peripheral (40-80\%) collisions at $\sqrt{s_{NN}}$ = 7.7, 11.5 and 39 GeV are compared to the UrQMD model prediction. A sign change in proton slope going from central to peripheral collisions  is observed. UrQMD is qualitatively consistent with the data  but does not predict the right magnitude at these energies. AMPT(default) model calculation is similar to UrQMD results. 
\begin{figure}
\begin{center}
\includegraphics[scale=0.54]{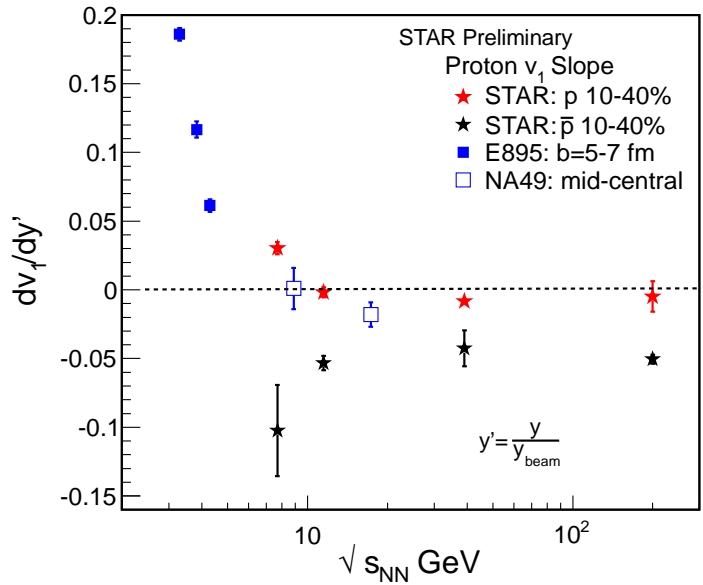}
\caption{Proton and antiproton slope $ dv_1/dy' $ at mid-rapidity as a function of beam energy. Antiproton data are not available from NA49 and E895.}
\label{fig4}
\end{center}
\end{figure}

In Fig.~\ref{fig4}, proton and antiproton slope $F = dv_1/dy'$ around midrapidity is plotted as a function of collision energy, where $y'$ is the scaled rapidity, defined as $y' = y/y_{\rm beam}$. Values for the slope of $v_1(y')$ are extracted via a polynomial fit of the form $Fy' + Cy'^3$. The proton slope decreases rapidly with increasing energy, changes its sign to negative  between 7.7 and 11.5 GeV, and remains small and negative up to 200 GeV.  In contrast, the antiproton slope, which was not reported by the NA49 or E895 collaborations, remains always negative in the measured $\sqrt{s_{NN}}$ range from 7.7 to 200 GeV.

 \section{Summary and Outlook}
In this proceedings, we present STAR results for directed flow as a function of transverse  momentum, pseudorapidity and centrality for Au+Au collisions at  $\sqrt{s_{NN}}$ = 7.7, 11.5 and  39 GeV. 
Our findings  demonstrate that  $v_1(\eta/y_{\rm beam})$  shows a beam energy scaling behavior, though not perfect, that has already been established at higher RHIC energies. Difference in 
directed flow of protons and antiprotons is observed. For midcentral collisions (10-40\% ), the $\pi^\pm$, $K^\pm$,  and antiprotons have a negative
   $dv_{1} /dy'$ at  mid rapidity, but proton $dv_{1}/dy'$ at 7.7 GeV becomes positive. The proton $v_{1}(y')$ slope decreases rapidly with increasing energy up to  7.7 GeV.  Its sign changes to 
   negative between 7.7 and 11.5 GeV, and remains small and negative at 11.5, 17.3, 39 and 200 GeV. However antiproton slope is always negative in the measured range from 7.7 GeV to 200 
   GeV.  Tested models do not predict the right magnitude of directed flow but UrQMD and AMPT (default) qualitatively follow some of the features shown by the data. The beam energy scan at 
   RHIC will continue in the future to map in more detail the interesting energy range between 11.5 and 39 GeV. Hence results from 19.6 GeV and 27 GeV, which were collected in 2011, will 
   provide further  important information to the observed beam energy dependence of the directed flow.

\end{document}